\documentclass[12pt] {article}		
\usepackage{epsfig}

\newlength{\figurewidth}
\normalsize
\newcommand{\be}{\begin{equation}}
\newcommand{\ee}{\end{equation}}
\newcommand{\bea}{\begin{eqnarray}}
\newcommand{\eea}{\end{eqnarray}}
\newcommand{\gsim}{\hbox{ \raise3pt\hbox to 0pt{$>$}\raise-3pt\hbox{$\sim$} }}
\newcommand{\lsim}{\hbox{ \raise3pt\hbox to 0pt{$<$}\raise-3pt\hbox{$\sim$} }}
\newcommand{\mathbold}[1]{\mbox{\boldmath $\bf#1$}}
\newcommand{\figdir}{.}
\begin{document}

\begin{titlepage}
\hfill

\vskip 2.0cm
\begin{center}
\large \bf
A Handy Tool for History Keeping of Geant4 Tracks\\
-- J4HistoryKeeper --
\end{center}


\vskip 1.5cm

\begin{center}
\large 
$Sumie$ $Yamamoto^a$, 
$Keisuke$ $Fujii^{b}$\footnote{
Corresponding author.\\
E-Mail address: keisuke.fujii@kek.jp\\
TEL: +8-29-864-5373\\
FAX: +8-29-864-2580
}, and 
$Akiya$ $Miyamoto^b$,
\end{center}

\vskip 0.7cm

\begin{center}
$^a$ School of High Energy Accelerator Science, 
 The Graduate University for Advanced Studies (Sokendai),
 Tsukuba, 305-0801, Japan\\ 
$^b$ High Energy Accelerator Research Organization (KEK),
 Tsukuba, 305-0801, Japan\\ 
\end{center}

\vskip 1cm

\begin{abstract}

The Particle Flow Analysis (PFA) is currently under intense studies
as the most promising way to achieve precision jet energy measurements
required at the future linear $e^+e^-$ collider.
In order to optimize detector configurations and to tune up the PFA
it is crucial to identify factors that limit the PFA performance 
and clarify the fundamental limits on the jet energy resolution
that remain even with the perfect PFA and an infinitely granular calorimeter.
This necessitates a tool to connect each calorimeter hit in
particle showers to its parent charged track, if any,
and eventually all the way back to
its corresponding primary particle, while identifying
possible interactions and decays along the way.
In order to realize this with a realistic memory space,
we have developed a set of C++ classes that
facilitates history keeping of particle tracks within the framework of Geant4.
This software tool, hereafter called J4HistoryKeeper, 
comes in handy in particular when one needs to stop
this history keeping for memory space economy 
at multiple geometrical boundaries beyond which 
a particle shower is expected to start.
In this paper this software tool is described and applied to a generic
detector model to demonstrate its functionality.
\end{abstract}

\vfil

\end{titlepage}


\section{Introduction}

The experiments at the International Linear Collider\cite{Ref:ILC}
will open up a novel possibility to reconstruct all the final states
in terms of fundamental particles (leptons, quarks, and gauge bosons)
as viewing their underlying Feynman diagrams.
This involves identification of heavy unstable particles
such as $W$, $Z$, $t$, and even yet undiscovered new
particles such as $H$ through jet invariant-mass measurements.
The goal is thus to achieve an jet invariant-mass resolution
comparable to the natural width of $W$ or $Z$\cite{Ref:JLC-I}. 
High resolution jet energy measurements will thus be crucial,
necessitating high resolution tracking and calorimetry
as well as an algorithm to make full use of available information
from them.
With a currently envisaged tracking system\cite{Ref:ILC} that aims at a momentum
resolution of $\sigma_{p_T} / p_T = 5 \times 10^{-5} p_T [{\rm GeV/c}]$ or better,
tracker information will be much more accurate than that from calorimetry
for charged particles.
This implies that the best attainable jet energy resolution should be achieved 
when we use the tracker information for charged particles and
the calorimeter information only for neutral particles.
This requires separation of calorimeter clusters due to
individual particles and, in the case of charged particle clusters,
their one-to-one matching to the corresponding tracks detected in the tracking system.
This is the so-called Particle Flow Analysis (PFA) currently under intense studies\cite{Ref:PFA}.

For the PFA, it is hence desirable to have a highly granular calorimeter that
allows separation of clusters due to a densely packed jet of particles. 
In practice the performance of the PFA depends not only on the hardware design
of the detector system consisting of the tracker and the calorimeter
but also on a particular algorithm one
employs to separate calorimeter signals due to neutral particles from
those due to charged particles.  
Various realistic algorithms are currently being tested by various groups\cite{Ref:PFA}.

For the optimization of detector configurations and the PFA algorithm, 
it is crucial to identify factors that limit the PFA performance 
and clarify the fundamental limits on the jet energy resolution
that remain even with an infinitely granular calorimeter and
an ideal algorithm to achieve perfect track-to-cluster matching.
We hence need a tool to connect each calorimeter hit in
particle showers to its parent charged track, if any,
and eventually all the way back to
its corresponding primary particle, while identifying
possible interactions and decays that might have taken place
along the way.
In order to achieve this history keeping with a reasonable memory size,
we need an algorithm to effectively achieve infinite calorimeter segmentation 
independently of the physical size of its readout cells
as well as a mechanism to stop history keeping
at various geometrical boundaries beyond which
particle showering is expected.

We have developed a set of C++ classes that realize such a functionality 
within the framework of Geant4.
We call this software tool J4HistoryKeeper hereafter,
since it is the name of the central class of the package.
Although J4HistoryKeeper was designed primarily for PFA studies, 
it has wider applications.
It comes in handy in particular when one needs to stop
history keeping for memory space economy 
at multiple user-registered geometrical boundaries.
The software tool was implemented in a Geant4\cite{Ref:geant4toolkit} based Monte Carlo simulator 
called JUPITER\cite{Ref:acfa,Ref:Jupiter} and has been used successfully for PFA studies, 
together with a smearing and reconstruction package called SATELLITES\cite{Ref:Jupiter},
running under a modular analysis frame work called JSF\cite{Ref:JSF}, 
both of which are based on ROOT\cite{Ref:root}.
The source code of a demo package of J4HistoryKeeper with slimmed up versions
of JUPITER and SATELLITES is available 
from our "J4HistoryKeeper Sample Code Page"\cite{Ref:j4hkeeper}.

In the following sections, we first elucidate the concept of {\it Cheated PFA},
which can be considered as the primary application that J4HistoryKeeper is designed for.
We then describe the software tool to keep history of particle tracks
(Geant4 tracks) traced through a detector in Geant4
with emphasis put on its design philosophy.
The subsequent  section is devoted to its usage with a sample
application to a generic ILC detector model to demonstrate
its functionality.

\section{Concept of Cheated PFA}

In principle Monte Carlo simulations allow us to use so-called
Monte-Carlo truth and enable us to unambiguously separate
calorimeter hits due to different incident particles,
thereby performing perfect clustering.
By linking so-formed calorimeter clusters to corresponding 
charged particle tracks in the tracking system again using
Monte-Carlo truth, we can achieve the situation with the perfect PFA.
We call this Cheated PFA (CPFA) since it involves cheating
by using Monte-Carlo truth, which is impossible in practice.
The concept of the CPFA is detailed in this section 
so as to clarify the philosophy behind the design of the software tool.


\subsection{Perfect Clustering and Perfect Track-Cluster Matching}

\begin{figure}[ht]
\begin{center}\begin{minipage}{\figurewidth}
\centerline{
\epsfig{file=\figdir/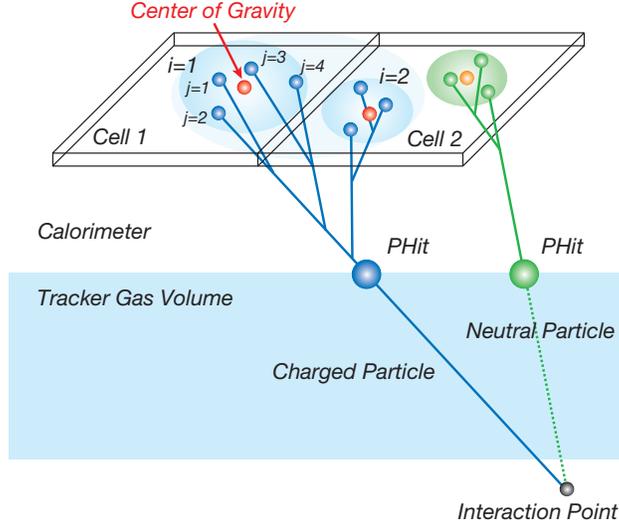,height=7cm}}
\caption[Fig:cpfa]{\label{Fig:cpfa} \small \it
Schematics showing the cheated PFA concept. 
Only two cells in a single sampling layer of the calorimeter are shown 
to simplify the picture though in practice much more cells are expected to be hit
over different sampling layers.
}
\end{minipage}\end{center}
\end{figure}

For the CPFA, 
the history of Geant4 tracks should be kept on a track-by-track basis
starting from a primary track at the interaction point.
The history of all the secondary tracks together with the original one
should be recorded until they hit  any one of pre-registered boundaries
beyond which particles may start showering.
At such a boundary we create a virtual hit called {\tt PHit}.
Calorimeter hits by Geant4 tracks in a particle shower 
will then be tagged with this {\tt PHit}. 
By collecting all the calorimeter hits with the same {\tt PHit}
we can hence form a calorimeter cluster without any confusion
(see Fig.\ref{Fig:cpfa}).
%
%

Since the {\tt PHit} carries the information of its parent track,
one-to-one matching between the calorimeter cluster and
its corresponding charged particle track in the tracking system
is possible.
Once matched, we just lock the calorimeter cluster as linked to a
charged track and just use the tracker information.
Calorimeter clusters with no matching charged tracks are hereafter
called neutral PFOs, while all the charged tracks are called
charged PFOs regardless of
whether there are corresponding calorimeter clusters or not.

It is also important to record the mother-daughter correspondence 
for particles decayed in a tracking volume
so as to estimate their effects on the PFA performance.
The mother-daughter correspondence is book-kept together with
the other information on the daughter track such as its particle ID, 
position, and momentum,
in a so-called {\tt BreakPoint} object which is created at the beginning
of each track.
The information stored in the {\tt BreakPoint} objects will be used
to follow particle decays observed as kinks or V$^0$s 
in the tracking volume and to
assign correct particle masses to charged PFOs.
This bookkeeping comprises the major role of the history keeper.

\subsection{Infinite Calorimeter Segmentation}

For a realistic calorimeter design, 
the granularity of the calorimeter, or equivalently the cell size, is
finite and hence the signals created by shower particles
stemming from different parent particles sometimes merge into a single hit
degrading the cluster separation capability.
In order to investigate to what extent this limits the PFA performance,
we need to know the performance expected for perfect cluster separation.
It is, however, impracticable to implement infinitely fine segmentation
even in a Monte Carlo detector simulator
because of memory space limitation.
In order to overcome this drawback,
we exploit the following trick.

In each hit cell, say cell $i$, 
we separately store the energy sum of hits originating from the same {\tt PHit}:
\begin{eqnarray}
E^{c}_i & = & \sum_{j} \, E_{i, j}  
\nonumber
\end{eqnarray}
and their center of gravity: 
\begin{eqnarray}
\mathbold{x}^{c}_i & = & \sum_{j} \, E_{i, j} \, \mathbold{x}_{i, j}  \, / \, E^{c}_i  ,
\nonumber
\end{eqnarray}
instead of using the cell center as the hit position.
In the above expression $E_{i, j}$ and $\mathbold{x}_{i, j}$ are the energy deposit 
and the position of $j$-th hit in cell $i$ with the same {\tt PHit}.
Denoting the total energy of the cluster originating from the same {\tt PHit} by
\begin{eqnarray}
E^{c} & = & \sum_{i} \, E^{c}_{i}   ,
\nonumber
\end{eqnarray}
we can then calculate its cluster center as
\begin{eqnarray}
\mathbold{x}^{c} & = & \sum_{i} \, E^{c}_{i} \, \mathbold{x}^{c}_{i} \, / \, E^{c}  
~~ = ~~ \sum_{i}  \sum_{j} \, E_{i, j} \, \mathbold{x}_{i, j} \, / \, \sum_{i} \, \sum_{j} \, E_{i, j}   ,
\nonumber
\end{eqnarray}
showing that the center of gravity calculated this way 
precisely coincides the one would-be obtained when the segmentation is infinitely fine.
It should be also emphasized here that 
hits from different {\tt PHit}s make multiple centers of gravity in the same cell,
which can be later separated even though they are in the same cell,
thereby realizing the infinite segmentation in effect.

\section{Tool Design}

Before coding our tool for history keeping, we set the following guideline
to fulfill the required functionality discussed in the last section:
a) there must be a versatile mechanism to register user-defined physical volumes 
	whose specified boundaries can be used to define a {\tt PHit} that marks the 
	source point of a particle shower,
b)
whether a track is allowed to create a {\tt PHit} or not depends on 
	whether the track originates from any pre-created {\tt PHit} or not,  
	which should be checked on a track-by-track basis at the beginning of its tracking, 
c)
 the history keeping is to be done on the track-by-track basis by creating
	a {\tt BreakPoint} object at the beginning of each track if there is no {\tt PHit}
	from which the track stems, and
d) the history keeping should be realized making maximum use of 
	existing Geant4 facilities within the framework of JUPITER,
e) JUPITER should produce Monte-Carlo truths (i.e. exact hits) and their smearing
	should be done later in SATELLITES as needed.

\begin{figure}[p]
\centerline{
\epsfig{file=\figdir/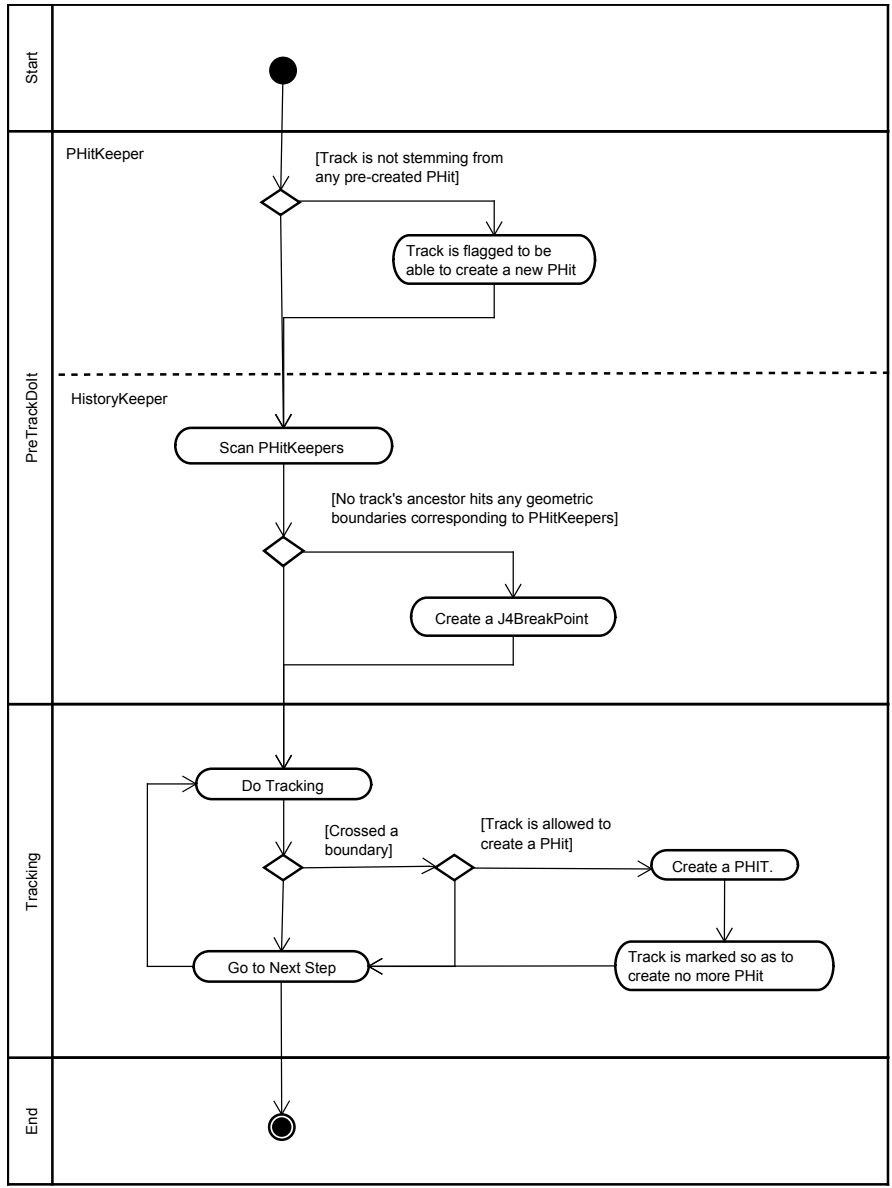,width=12cm}}
\caption[Fig:ActivityPlot]{\label{Fig:ActivityPlot} \small \it
The flow diagram for Geant4 tracking related to J4HistoryKeeper.
}
\end{figure}

\begin{flushleft}
The following is a sketch of the tool design we adopted according to the guideline:
\end{flushleft}
\begin{enumerate}
\item The history keeping is to be done on the track-by-track basis using {\tt J4TrackingAction}
	that inherits from {\tt G4UserTrackingAction}. 
	Its {\tt PreUserTrackingAction} method is hence called
	at the beginning of a new track.
	The {\tt PreUserTracingAction} method serially invokes {\tt PreTrackDoIt} method of
	each offspring of {\tt J4VSubTrackingAction} pre-registered to the {\tt J4TrackingAction}
	object. 
	Likewise, its {\tt Clear} method serially invokes {\tt Clear} method of
	individual offsprings of {\tt J4VSubTrackingAction}.  
\item {\tt J4VSubTrackingAction} is an abstract class that serves as a base class for 
	user-defined sub-actions 
	taken by {\tt J4TrackingAction} thereby extending the 
	{\tt G4UserTrackingAction} functionality.
	It has a method called {\tt Clear} to reset the object state. 
\item {\tt J4HistoryKeeper} is implemented as a derived class from
	{\tt J4VSubTrackingAction}
	and, in its {\tt PreTrackDoIt} method,
	scans through a collection of pre-registered {\tt J4PHitKeeper} objects corresponding
	to a collection of bounding surfaces.
	It then  creates a {\tt J4BreakPoint} object 
	if none of them has been
	hit by any ancestors of the new track, 
\item {\tt J4PHitKeeper} also inherits from {\tt J4VSubTrackingAction}.
	Its {\tt PreTrackDoIT} method  checks if this new track is 
	stemming from any pre-created {\tt PHit}, and, if not, resets its sate
	to allow creation of a new {\tt PHit}.
	When its corresponding boundary is hit by the current track,
	a {\tt PHit} object is created, if it is allowed,
	to tag subsequent daughter tracks possibly created in a shower.
\end{enumerate}
The flow of tracking related to J4HistoryKeeper is shown in Fig.~\ref{Fig:ActivityPlot}.

\subsection{Extension of  G4UserTrackingAction}

The {\tt G4UserTrackingAction} class provides one with a handy tool 
to perform a user-defined action on a track-by-track basis.
In its original form, however, it allows only a single action.
In order to extend its functionality to accept multiple user-defined actions,
we have introduced the concept of {\tt SubTrackingAction} as sketched above.
\begin{flushleft}
\underline{{\tt J4VSubTrackingAction}}
\end{flushleft}
The abstract base class {\tt J4VSubTrackingAction} has the following methods: 
\begin{description}
	\item {\tt void PreTrackDoIt(const G4Track *aTrack = 0) = 0~}\\
		which is pure virtual and to be implemented by
		its derived class to take a sub-tracking action for the given track.
	\item {\tt void Clear()~}\\
		which does nothing, and to be overridden in the derived class as needed.
\end{description}
This class just specifies the interface and requires its users to implement
the methods listed above.

\begin{figure}[ht]
\centerline{
\epsfig{file=\figdir/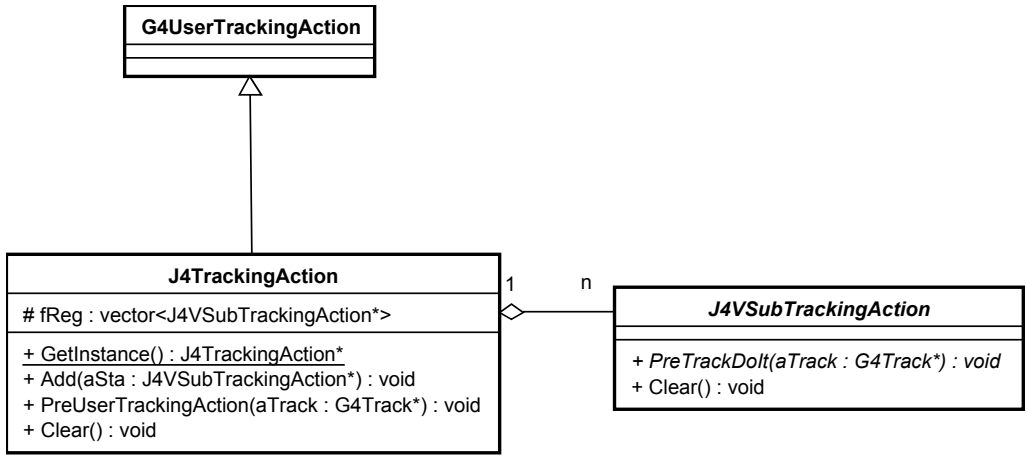,width=14cm}}
\caption[Fig:J4TrackingAction]{\label{Fig:J4TrackingAction} \small \it
Class diagram for {\tt J4TrackingAction} and {\tt J4VSubTrackingAction}.}
\end{figure}

\begin{flushleft}
\underline{{\tt J4TrackingAction}}
\end{flushleft}
The {\tt J4TrackingAction} class is a singleton inheriting from {\tt G4UserTrackingAction}.
It has, among others, an STL vector as a data member to store pointers to objects 
derived from the {\tt J4VSubTrackingAction} class.
Its major methods include the following:
\begin{description}
	\item {\tt static J4TrackingAction *GetInstance()~}\\
		which returns the pointer to the single instance of {\tt J4TrackingAction}.
	\item {\tt void Add(J4VSubTrackingAction *aSta)~}\\
		which registers a user-defined object derived from {\tt J4VSubTrackingAction}.
		When {\tt *aSta} has already been registered, the pre-registered one is erased
		and the new entry is appended.
	\item {\tt void PreUserTrackingAction(const G4Track *aTrack)~}\\
		which loops over the registered offsprings of {\tt J4VSubTrackingAction}
		and invokes their {\tt PreTrackDoIt} methods.
	\item {\tt void Clear()~}\\
		which loops over the registered offsprings of {\tt J4VSubTrackingAction}
		and invokes their {\tt Clear} methods.
\end{description}
The class diagram for {\tt J4TrackingAction} and 
{\tt J4VSubTrackingAction} is shown in Fig.~\ref{Fig:J4TrackingAction}.

\subsection{P-Hits and P-Hit Keeper}

\begin{figure}[ht]
\centerline{
\epsfig{file=\figdir/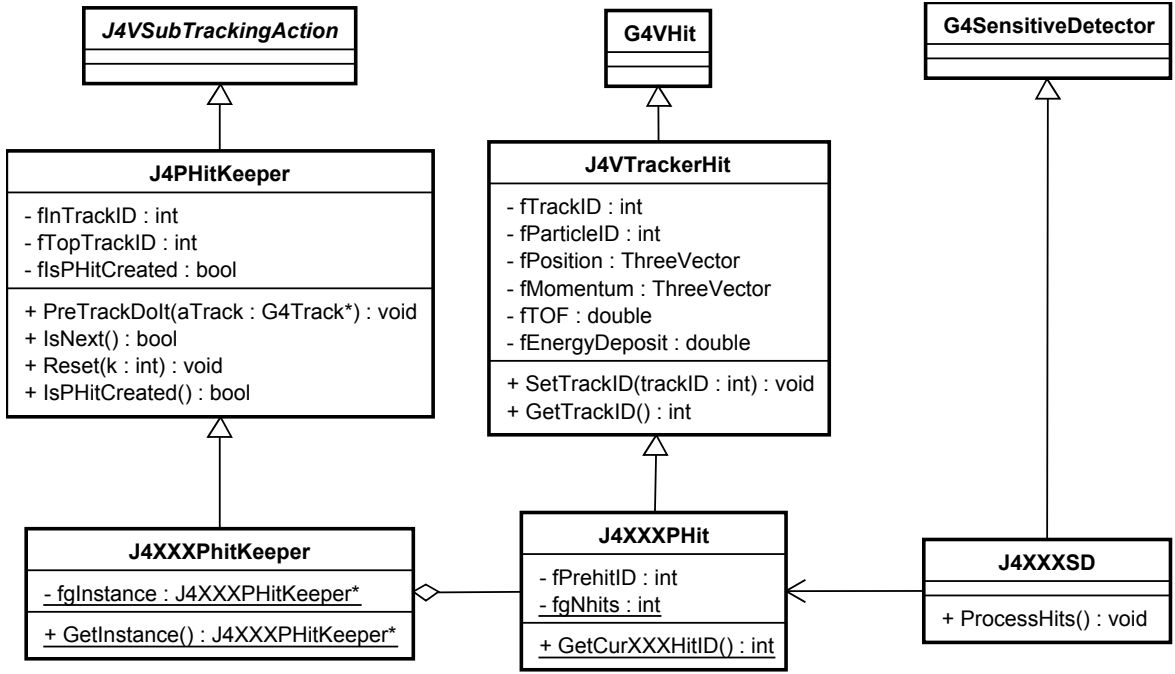,width=16cm}}
\caption[Fig:J4PHitKeeper]{\label{Fig:J4PHitKeeper} \small \it
The class diagram for {\tt J4PHitKeeper}, {\tt J4VTrackerHit}, and related classes.}
\end{figure}

{\tt PHit} is a generic name for a {\tt Pre-Hit} or a {\tt Post-Hit}, which
stands for a virtual hit created
on a boundary of a {\tt G4PhysicalVolume}
beyond which particle showering is expected.
The {\tt PHit} creation is done in the user-overridden {\tt ProcessHits} method
of a user-defined virtual detector derived from
{\tt G4SensitiveDetector} corresponding to the physical volume.
Notice that {\tt PHit}s are created for all kinds of particles, even neutrinos, that pass through
the boundary.
One {\tt PHit} class is defined inheriting from the {\tt J4VTrackerHit} class
for each such boundary.
The {\tt J4VTrackerHit} class carries basic track hit information such as
track ID, particle ID, position, momentum, TOF, energy deposit, etc. and 
setters and getters to access them.
An individual {\tt PHit} class has a data member to store {\tt PHit} ID and
a static data member to store the current {\tt PHit} ID, 
which can be retrieved by a static method to mark calorimeter hits as needed.
The class diagram for {\tt J4VTrackerHit}, {\tt J4PHitKeeper}, 
and related classes is shown in Fig.~\ref{Fig:J4PHitKeeper}.

\begin{flushleft}
\underline{{\tt J4PHitKeeper}}
\end{flushleft}
The {\tt J4PHitKeeper} class inherits from {\tt J4VSubTrackingAction}.
It serves as a base class for a {\tt PHitKeeper} class defined for
an individual {\tt PHit} class corresponding to a boundary beyond which
particle showering is expected.
The {\tt J4PHitKeeper} class has data members to store i) the current incident track ID 
({\tt fInTrackID}) that is expected to create or has already created a {\tt PHit}, 
ii) the track ID ({\tt fTopTrackID}) of the next track to be processed, if any,  
after the offsprings from the {\tt PHit} are exhausted,
and iii) a flag ({\tt fIsPHitCreated}) to tell whether a {\tt PHit} has been created or not.
The major methods of {\tt J4PHitKeeper} are listed below:
\begin{description}
	\item {\tt void PreTrackDoIt(const G4Track *)~}\\
		implements the corresponding base class pure virtual method
		so as to reset {\tt fInTrackID} and {\tt fTopTrackID} to {\tt std::numeric\_limits<int>::max()}
		and {\tt fIsPHitCreated} to {\tt false} upon encountering a new track
		which has a track ID smaller than {\tt fTopTrackID}.
	\item {\tt G4bool IsNext()~}\\
		returns {\tt false} if a {\tt PHit} has already been created.
		If not, it updates {\tt fInTrackID} and {\tt fTopTrackID} and
		returns {\tt true} to tell the caller (the {\tt ProcessHits} method of the 
		sensitive detector defining the virtual boundary) 
		that a new {\tt PHit} is to be created.
	\item {\tt void Reset(G4int k = std::numeric\_limits<int>::max())~}\\
		resets {\tt fInTrackID} and {\tt fTopTrackID} to {\tt k}.
	\item {\tt G4bool IsPHitCreated()~}\\
		returns {\tt fIsPHitCreated}, which is {\tt true} if a {\tt PHit} has been created,
		and {\tt false} otherwise.
\end{description}
The algorithm of {\tt J4PHitKeepr} heavily depends on Geant4's default track stacking scheme,
which is worth explaining here for readers unfamiliar with it.
By default Geant4 uses two types of track stacks, a Primary Stack ({\it PS}) 
and a Secondary Stack ({\it SS}).

At the beginning of each event, primary particles $1, \cdots, n$ are pushed into {\it PS}.
According to the "last in first out" rule, the top entry, track $n$, is popped out for tracking.
Notice that there remains $n-1$ tracks in {\it PS} at this point.
All the secondary particles produced while track $n$ is being processed are
pushed into {\it SS}.
Let us assume that there will be $m$ secondary particles stacked into {\it SS} 
by the time track $n$ is disposed of.
All of these $m$ secondary particles in {\it SS} are moved to {\it PS} upon the death
of track $n$ and numbered serially as tracks $n+1, \cdots, n+m$.
Notice that there are $n+m-1$ tracks in {\it PS} at this point since
track $n$ has been popped out and disposed of.

The key point is to bookmark the secondary track which is to be created just
after the creation of a {\tt PHit} by the track which has been being processed,
track $n$ in the present case.
The track ID with the bookmark will be ${\tt fTopTrackID} = n+k'+1$ 
where $k' (\le m)$ is the number of secondary particles in {\it SS} at the time of
the {\tt PHit} creation.
Further {\tt PHit} creation is to be forbidden until it becomes necessary.

The top of the stack, track $n+m$, is popped out and to be processed as before.
Track $n+m$ will produce further $m'$ secondary particles to be pushed into
{\it PS} upon its death and to be numbered as tracks $n+m+1, \cdots, n+m+m'$.

This procedure is repeated and after some time 
all the secondary particles originating from the track created the last
{\tt PHit} will be disposed of and
the next track to be popped out from {\it PS} will have a track ID that is
smaller than that of the last bookmarked one, {\tt fTopTrackID}.
This signals a new incident track which is allowed to create a new {\tt PHit}.
By repeating this procedure until all the tracks in {\it PS} are exhausted,
we can mark all the calorimeter hits with corresponding {\tt PHit}s.

\subsection{Break Points and History Keeper}

\begin{figure}[ht]
\centerline{
\epsfig{file=\figdir/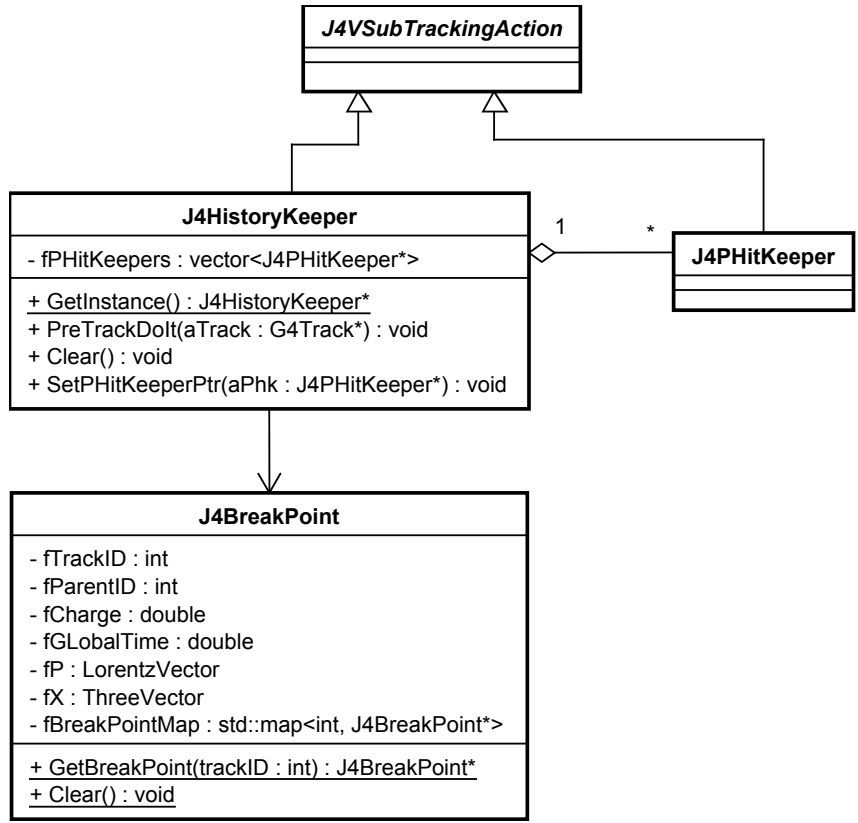,width=12cm}}
\caption[Fig:J4HistroyKeeper]{\label{Fig:J4HistroyKeeper} \small \it
Class diagram for {\tt J4HistroyKeeper} and {\tt J4BreakPoint}.}
\end{figure}

The purpose of the history keeper is to allow us to trace back to
kink and V$^0$ particles that decay before entering calorimeters
so as to correctly link clusters to tracks.
As sketched above, the history keeper is implemented as a
{\tt J4VSubTrackingAction} so as to create
a {\tt J4BreakPoint} object for each new track until
a {\tt PHit} is created on any of the pre-registered boundaries
beyond which particle-showering is expected.
The class diagram for {\tt J4BreakPoint} and {\tt J4HistoryKeeper} is shown in 
Fig.~\ref{Fig:J4HistroyKeeper}.

\begin{flushleft}
\underline{{\tt J4BreakPoint}}
\end{flushleft}
The {\tt J4BreakPoint} class has data members to store
the information about a track at its starting position such as
track ID ({\tt fTrackID}), parent track ID ({\tt fParentID}), 
charge, particle ID,  time, position, 4-momentum, etc..
In addition it has a static data member called {\tt fgBreakPointMap},
which is an STL map that links track ID to a {\tt J4BreakPoint} object.
Besides the getters to these data members, 
{\tt J4BreakPoint} has the methods listed below:
\begin{description}
	\item {\tt static J4BreakPoint *GetBreakPoint(G4int trackID)~}\\
		returns the pointer to the {\tt J4BreakPoint} object
		corresponding {\tt trackID}.
	\item {\tt static void Clear()~}\\
		clears the track-to-break-point map.
\end{description}

\begin{flushleft}
\underline{{\tt J4HistoryKeeper}}
\end{flushleft}
The {\tt J4HistoryKeeper} class is a singleton that inherits from {\tt J4VSubTrackingAction}.
It has an STL vector ({\tt fPHitKeepers})
as a data member to store registered {\tt J4PHitKeeper}s that
correspond to boundaries beyond which particle-showering is expected.
As sketched above, 	it scans through these pre-registered {\tt J4PHitKeeper} objects
to make sure that none of them has a {\tt PHit}, 
and then creates a {\tt J4BreakPoint} object.
The major methods of {\tt J4HistoryKeeper} are listed below:
\begin{description}
	\item {\tt static J4HistoryKeeper *GetInstance()~}\\
		returns the pointer to the single instance of {\tt J4HistoryKeeper}.
	\item {\tt void PreTrackDoIt(const G4Track *)~}\\
		implements the corresponding base class pure virtual method.
		It scans through the pre-registered {\tt J4PHitKeeper} objects
		in {\tt fPHitKeepers} to make sure that none of them has a {\tt PHit}
		by calling their {\tt IsPHitCreated()} method.
		It then creates a {\tt J4BreakPoint} object.
	\item {\tt void Clear()~}\\
		calls {\tt J4BreakPoint::Clear()}.
	\item {\tt void SetPHitKeeperPtr(J4PHitKeeper *aPhk)~}\\
		pushes back the input {\tt J4PHitKeeper} pointer into {\tt fPHitKeepers}.
\end{description}

\begin{figure}[ht]
\centerline{
\epsfig{file=\figdir/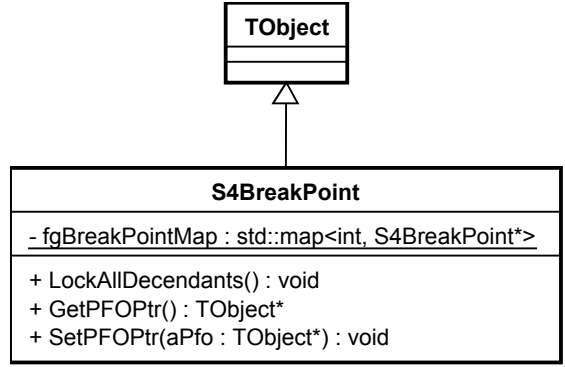,width=8cm}}
\caption[Fig:S4BreakPoint]{\label{Fig:S4BreakPoint} \small \it
Class diagram for {\tt S4BreakPoint}.}
\end{figure}

\begin{flushleft}
\underline{{\tt S4BreakPoint}}
\end{flushleft}
Upon the completion of Monte Carlo truth generation by JUPITER,
each {\tt J4BreakPoint} object is copied to its SATELLITE dual,
an {\tt S4BreakPoint} object.
The {\tt S4BreakPoint} object inherits from ROOT's {\tt TObjArray} and
stores pointers to its daughter {\tt S4BreakPoint}s, if any.
It has additional methods such as
\begin{description}
	\item {\tt void LockAllDescendants()~}\\
		which flags all of its descendants as locked. This functionality proves handy to
		avoid double counting of energies.
	\item {\tt TObject *GetPFOPtr()~}\\
		which returns the pointer to its corresponding Particle Flow Object (PFO), if any.\\
	\item {\tt void SetPFOPtr(TObject *aPfo)~}\\
		which is the setter corresponding to {\tt GetPFOPtr} to be
		invoked from a PFO maker.
\end{description}
The class diagram for {\tt S4BreakPoint} is shown in Fig.~\ref{Fig:S4BreakPoint}.

\section{Tool Usage}

What a tool user has to do for the history keeping is as follows:
\begin{itemize}
\item Inheriting {\tt G4SensitiveDetector},
	create a sensitive detector class, say {\tt J4XXXSD}, 
	that corresponds to 
	a boundary on which a {\tt PHit} object ({\tt J4XXXPHit}) is to be created 
	for each particle that is expected to produce a shower
	beyond that boundary.
\item Inheriting {\tt J4PHitKeeper},
	create a {\tt J4XXXPHitKeeper} as a singleton to book-keep {\tt J4XXXPHit}s.
\item In {\tt J4XXXSD}'s constructor, do
\begin{verbatim}
	  J4XXXPHitKeeper *aPhk = J4XXXPHitKeeper::GetInstance();
	  J4TrackingAction::GetInstance()->Add(aPhk);
	  J4TrackingAction::GetInstance()->Add(J4HistoryKeeper::GetInstance());
	  J4HistoryKeeper::GetInstance()->SetPHitKeeperPtr(aPhk);
\end{verbatim}
	in order to register the {\tt J4XXXPHitKeeper} to {\tt J4TrackingAction} and to
	{\tt J4HistoryKeeper}.
\item In {\tt J4XXXSD}'s {\tt ProcessHits($\cdots$)} method, do 
\begin{verbatim}
	  if (J4XXXPHitKeeper::GetInstance()->IsNext()) {
		  // create and store a J4XXXPHit object 
	  }
\end{verbatim}
\item In {\tt ProcessHits($\cdots$)} of each calorimeter sensitive detector, which
	usually corresponds to a single calorimeter cell, 
	store the centers of gravity and energy deposits of particles
	from different {\tt PHit}s as different calorimeter hits even in the same cell and
	mark them with the current {\tt PHit} ID obtainable from an
	appropriate {\tt J4PHitKeeper} object. 
\end{itemize}
This ensures the history keeping to be continued until any one of
the pre-registered boundaries is hit and beyond which the calorimeter 
hits are marked with the {\tt PHit} ID put to the {\tt PHit} created on that
boundary.

\begin{figure}[ht]
\begin{center}\begin{minipage}{\figurewidth}
\centerline{
\epsfig{file=\figdir/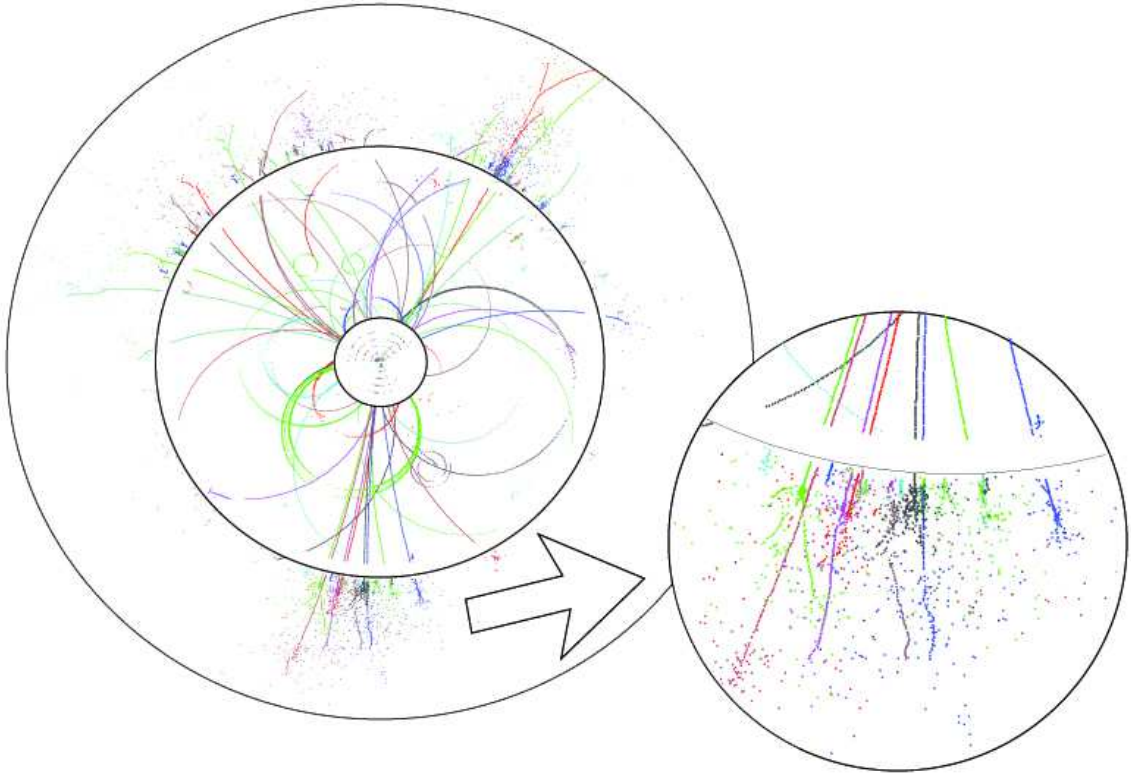,width=\textwidth}}
\caption[Fig:qqbar]{\label{Fig:qqbar} \small \it
A typical $e^{+}e^{-} \rightarrow q\bar{q}$ event with a hard gluon emission 
at the center of mass energy of 350 GeV, viewed from the beam direction.  
Matching calorimeter and tracker signals are identified by using the J4HisotryKeeper package
and painted with same colors. Three circles indicate, from outside to inside, 
the outer and the inner boundaries of the calorimeter, and the inner wall of 
the TPC. Points inside the TPC inner wall are signals by silicon trackers.
}
\end{minipage}\end{center}
\end{figure}

As an example of J4HistoryKeeper application, a typical multi-particle event 
detected with a generic $e^{+}e^{-}$ collider detector model is shown in Fig.~\ref{Fig:qqbar}.
It is an $e^{+}e^{-} \rightarrow q\bar{q}$ event at the center of mass energy 
of 350 GeV. Circles in the figure indicate, from outside to inside,
the outer and the inner boundaries of the barrel calorimeter, 
and the inner wall of the central tracker, a Time Projection Chamber (TPC), respectively.
The calorimeter inner radius of 210 cm sets the scale of the detector model.
Both the tracker and the calorimeter are placed in a solenoidal magnetic field of 3 Tesla.
Using the J4HistoryKeeper package, calorimeter signals  
are matched to their corresponding track signals and are painted with the same color.  
With J4HistoryKeeper, the centers of gravity of the calorimeter signals can be 
calculated as with infinite segmentation, 
which proves extremely useful in investigating the ultimate performance of 
the detector system with infinitely granular calorimetry.

%
%

\section{Conclusion}

We have developed a software tool, J4HistoryKeeper, for history keeping of Geant4 tracks.
J4HistoryKeeper records history of Geant4 tracks 
starting from the interaction point until they reach 
any of user-registered geometrical boundaries.
The tool allows us to record their positions, momenta, trackIDs, TOFs, etc.
at their birth points as well as at the user-registered boundaries.  
The flexible registration capability of the user-given boundaries
and a mechanism to achieve effectively-infinite segmentation
with a finite readout cell size comprise the core features of the package.
These features provide J4HistoryKeeper with wider applications though
it has been designed primarily for PFA studies.
The package comes in handy in any application where
history keeping is necessary but particle showering and 
consequently memory need explosion are expected
beyond multiple boundaries of a complicated detector geometry.
The tool has been used for the so-called Cheated PFA
to investigate limiting factors and ultimate performance
of jet energy measurements at the future linear $e^+e^-$ collider.

\section*{Acknowledgments}

The authors would like to thank T.~Yoshioka, H.~Ono, T.~Takeshita, 
and other members of the JLC-Software group
for useful discussions and helps.
This work was supported in part by the Creative Scientific Research 
Grant No. 18GS0202 of the Japan Society for Promotion of Science (JSPS)
and the JSPS Core University Program.


\begin{thebibliography}{99}

\bibitem{Ref:ILC}  http://www.linearcollider.org/ and references therein.

\bibitem{Ref:JLC-I} JLC group, KEK Report 92-16, December, 1992.

\bibitem{Ref:PFA} 
M.A.Thomson, arXiv:physics/060726;
V.Morgunov and A.Raspereza, arXiv:physics/0412108;
T.Yoshioka, ECONF C0508141:ALCPG1711,2005;
S.Magill and S.Kuhlmann, SLAC-PUB-12203,
in the Proceedings of 2005 International Linear Collider Workshop (LCWS 2005),
Stanford, California, 18-22 Mar 2005, pp 1015.

\bibitem{Ref:acfa}
ACFA Linear Collider Working Group, KEK Report 2001-11, August (2001),\\ 
http://acfahep.kek.jp/acfareport/.

\bibitem{Ref:Jupiter}
Proceedings of the APPI Winter Institute, KEK Proceedings 2002-08, July (2002).

\bibitem{Ref:geant4toolkit}
 http://wwwinfo.cern.ch/asd/geant4/G4UsersDocuments/\\UsersGuides/ForToolkitDeveloper/html/.
 
 \bibitem{Ref:JSF}
 http://acfahep.kek.jp/subg/sim/simtools/.
 
\bibitem{Ref:root}
http://root.cern.ch/.

\bibitem{Ref:j4hkeeper}
http://www-jlc.kek.jp/subg/offl/cpfa/.
 



\end{thebibliography}
\end{document}